\documentclass[aps,pre,twocolumn,showkeys,showpacs,floatfix]{revtex4-1}
\usepackage[OT4]{fontenc}
\usepackage[colorlinks=true,hyperfootnotes=true,breaklinks=true]{hyperref}
\usepackage{graphicx} 
\usepackage[figbotcap]{subfigure}
\usepackage[geometry]{ifsym}

\begin{document}

\title{Competing contact processes in~the~Watts--Strogatz network}

\author{Marcin Rybak}  
\author{Krzysztof Malarz}  
\homepage{http://home.agh.edu.pl/malarz/} 
\email{malarz@agh.edu.pl} 
\author{Krzysztof Ku{\l}akowski}
\email{kulakowski@fis.agh.edu.pl} 
\affiliation{\href{http://www.agh.edu.pl/}{AGH University of Science and Technology},
\href{http://www.pacs.agh.edu.pl/}{Faculty of Physics and Applied Computer Science},
al. Mickiewicza 30, 30-059 Krakow, Poland.}

\date{\today}

\begin{abstract}
We investigate two competing contact processes on a set of Watts--Strogatz networks with the clustering coefficient tuned by rewiring.
The base for network construction is one-dimensional chain of $N$ sites, where each site $i$ is directly linked to nodes labelled as $i\pm 1$ and $i\pm 2$.
So initially, each node has the same degree $k_i=4$.
The periodic boundary conditions are assumed as well.

For each node $i$ the links to sites $i+1$ and $i+2$ are rewired to two randomly selected nodes so far not-connected to node $i$.
An increase of the rewiring probability $q$ influences the nodes degree distribution and the network clusterization coefficient $\mathcal{C}$.
For given values of rewiring probability $q$ the set $\mathcal{N}(q)=\{\mathcal{N}_1, \mathcal{N}_2, \cdots, \mathcal{N}_M \}$ of $M$ networks is generated.

The network's nodes are decorated with spin-like variables $s_i\in\{S,D\}$. During simulation each $S$ node having a $D$-site in its neighbourhood converts this neighbour from $D$ to $S$ state. Conversely, a node in $D$ state having at least one neighbour also in state $D$-state converts all nearest-neighbours of this pair into $D$-state. The latter is realized with probability $p$.

We plot the dependence of the nodes $S$ final density $n_S^T$ on initial nodes $S$ fraction $n_S^0$.
Then, we construct the surface of the unstable fixed points in $(\mathcal{C}, p, n_S^0)$ space.
The system evolves more often toward $n_S^T=1$ for $(\mathcal{C}, p, n_S^0)$ points situated above this surface while starting simulation with $(\mathcal{C}, p, n_S^0)$ parameters situated below this surface leads system to $n_S^T=0$.
The points on this surface correspond to such value of initial fraction $n_S^*$ of $S$ nodes (for fixed values $\mathcal{C}$ and $p$) for which their final density is $n_S^T=\frac{1}{2}$.

\end{abstract}



\maketitle

\section{\label{sec-intro}Introduction}
In computational modeling, the contact processes (CPs) are dynamic systems on discrete media, where the time evolution of a local state towards survival or extinction of particles is determined by the state of the direct neighborhood of a lattice cell or of a network node. A simple realization is the voter model \cite{0}, where a particle creates another particle in its direct neighborhood. CPs have been introduced in 1974 as a toy model of spread of epidemic on a lattice \cite{1,2}. Since then, they evolved to a frame for models in different areas, from symbiotic interactions \cite{3} to population \cite{3a} or opinion dynamics \cite{4}. Yet, their important role is also to inspire theoretical considerations on non-equilibrium processes \cite{5,6,7}.  In both these roles, the spectrum of particular realizations of CPs has been remarkably enriched. In particular, the pair contact processes have been proposed in Ref.~\cite{8}; there, a pair of particles annihilate or create a neighbor particle.

The aim of this paper is to report our numerical results on competing CPs of two different kinds. As far as we know, this case has not been analyzed, with two our texts \cite{9d,9e} as an exception.  In literature, applications of competing CPs are of recent interest \cite{9a,9b,9c}; yet, in all these approaches the competing processes are of the same kind. In Ref.~\cite{9d}, the competition has been investigated between the voter model and the pair contact process without annihilation. The role of network topology has been analyzed by a comparison of results for the Watts--Strogatz network \cite{11} and the Erd{\H o}s--R\'enyi network \cite{12}, where the clustering coefficient $\mathcal{C}$ has been tuned in both networks. Our motivation in Ref.~\cite{9d} was to evaluate the efficiency of the pair contact process by balancing it with the voter model of controlled efficiency; the control was kept by tuning the probability $p$ of the one-node voter dynamics. The key result of Ref.~\cite{9d} was a phase 
diagram on the plane ($\mathcal{C},p$); below some critical line $p_1^c(\mathcal{C})$, the pair process dominates, while above another critical line $p_2^c(\mathcal{C})$, the voter dynamics prevails. Between these lines, {\em i.e.} for $p_1^c(\mathcal{C})<p<p_2^c(\mathcal{C})$, the time of relaxation was too long to get a definitive conclusion on the stability of this or that phase. Both critical lines have been found to depend on the network topology.

In paper \cite{9e}, preliminary results have been reported on the competition between the voter model dynamics and the Sznajd model dynamics \cite{10a,10b}. The latter algorithm bears some resemblance to the pair CP \cite{8}; yet, pairs do not annihilate there, and new particles are created at the whole neighborhood of the pair. As the result, we have got a slight dependence of the transition line in the phase diagram ($\mathcal{C},p$) on the initial conditions, {\em i.e}. on the initial percentage of nodes in the state activated by the voter dynamics. As we demonstrate below, this result appears to be generic. The work presented here is entirely devoted to the role of the initial conditions. This makes the problem more complex; the plane $(C,p)$ to construct the phase diagram is to be substituted by the three-dimensional space $(\mathcal{C},p,n_S^0)$.

The next section (\ref{sec-model}) is devoted to the model and to the details of our simulation procedure.
The section~\ref{sec-results} provides our numerical results.
In the last section (\ref{sec-summary}) we give a summary, supplemented by a note on a possible application of  the scheme presented here.

\section{\label{sec-model}Model}

\subsection{\label{sec-network}Network construction}

The simulations take place on networks similar to Watts--Strogatz networks \cite{11}.
The base for network construction is one-dimensional chain of $N$ sites, where each site $i$ is directly linked to nodes labeled as $i\pm 1$ and $i\pm 2$.
So initially, each node has the same degree $k_i=4$.
The periodic boundary conditions are assumed as well.

\begin{figure}
\begin{center}
\includegraphics[width=0.33\textwidth]{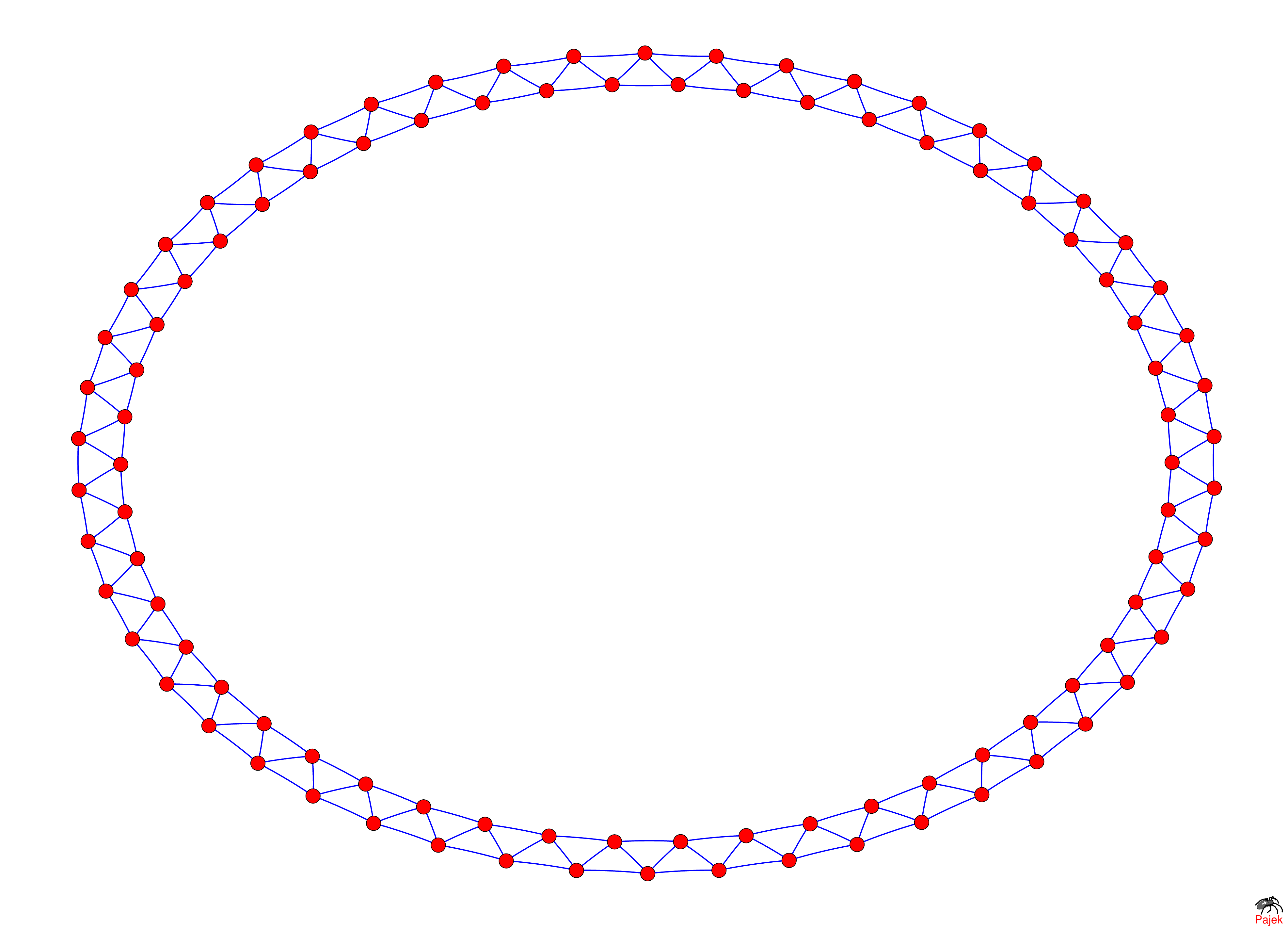}\\
\includegraphics[width=0.36\textwidth]{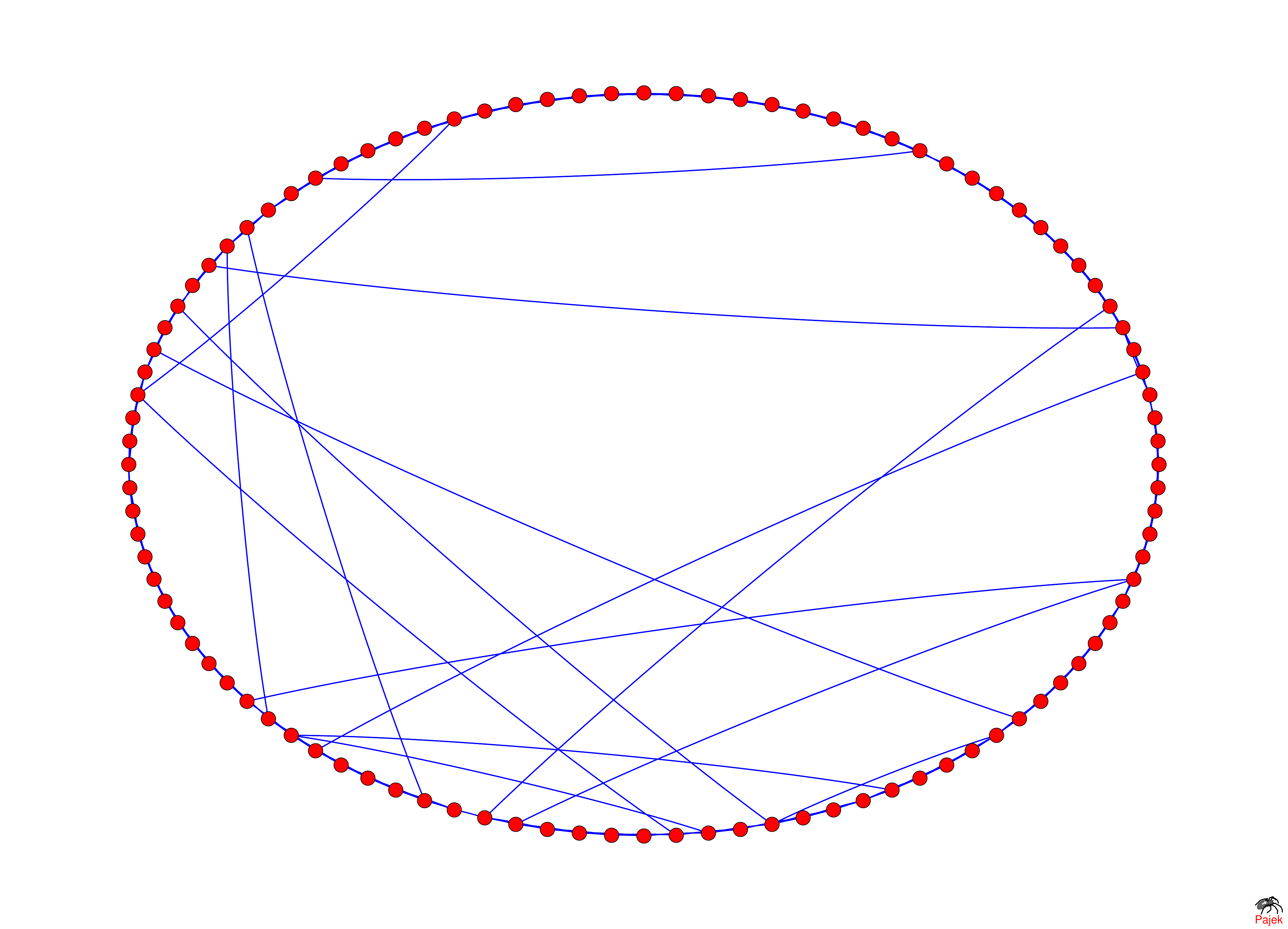}\\
\includegraphics[width=0.36\textwidth]{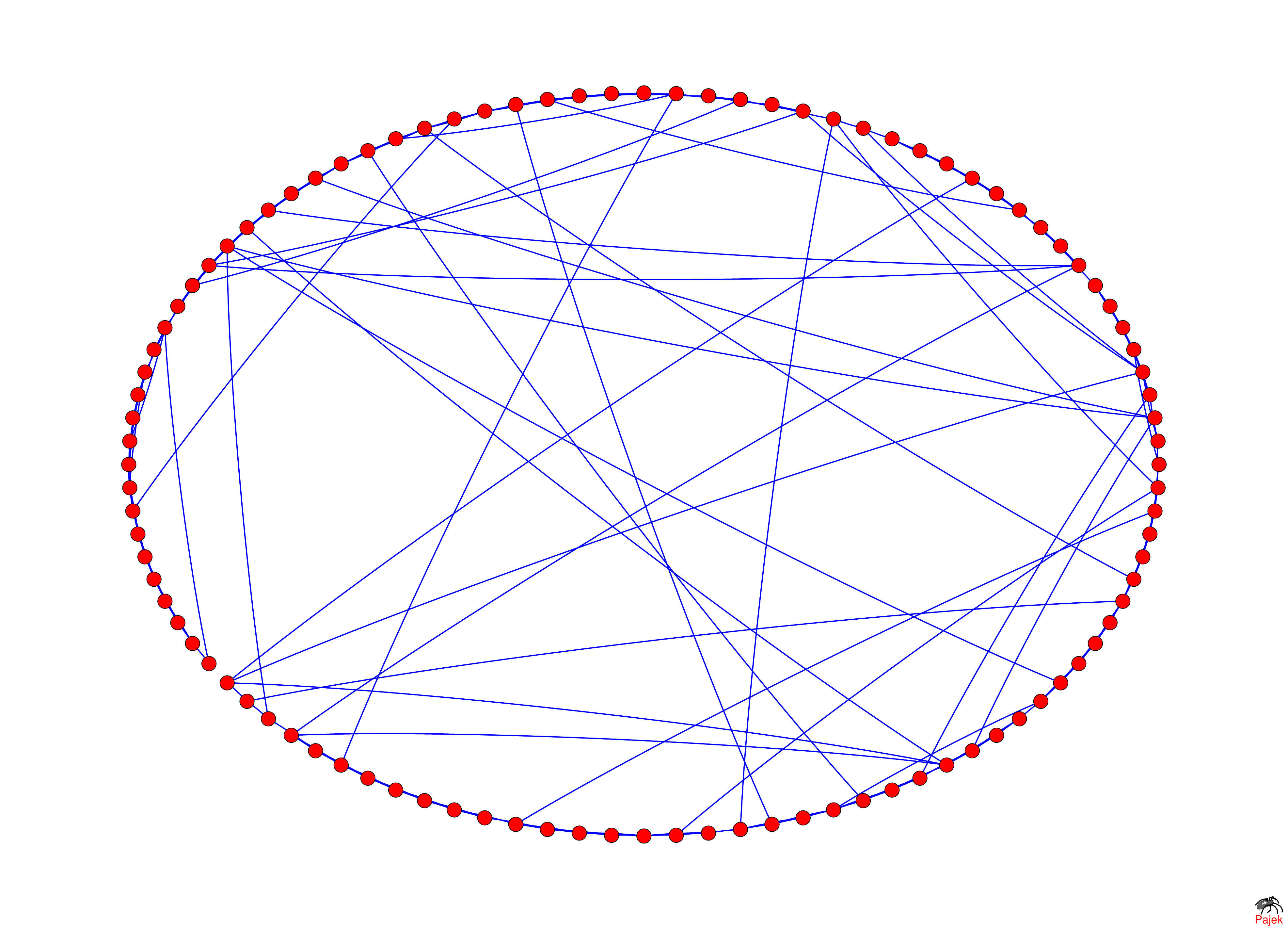}\\
\includegraphics[width=0.36\textwidth]{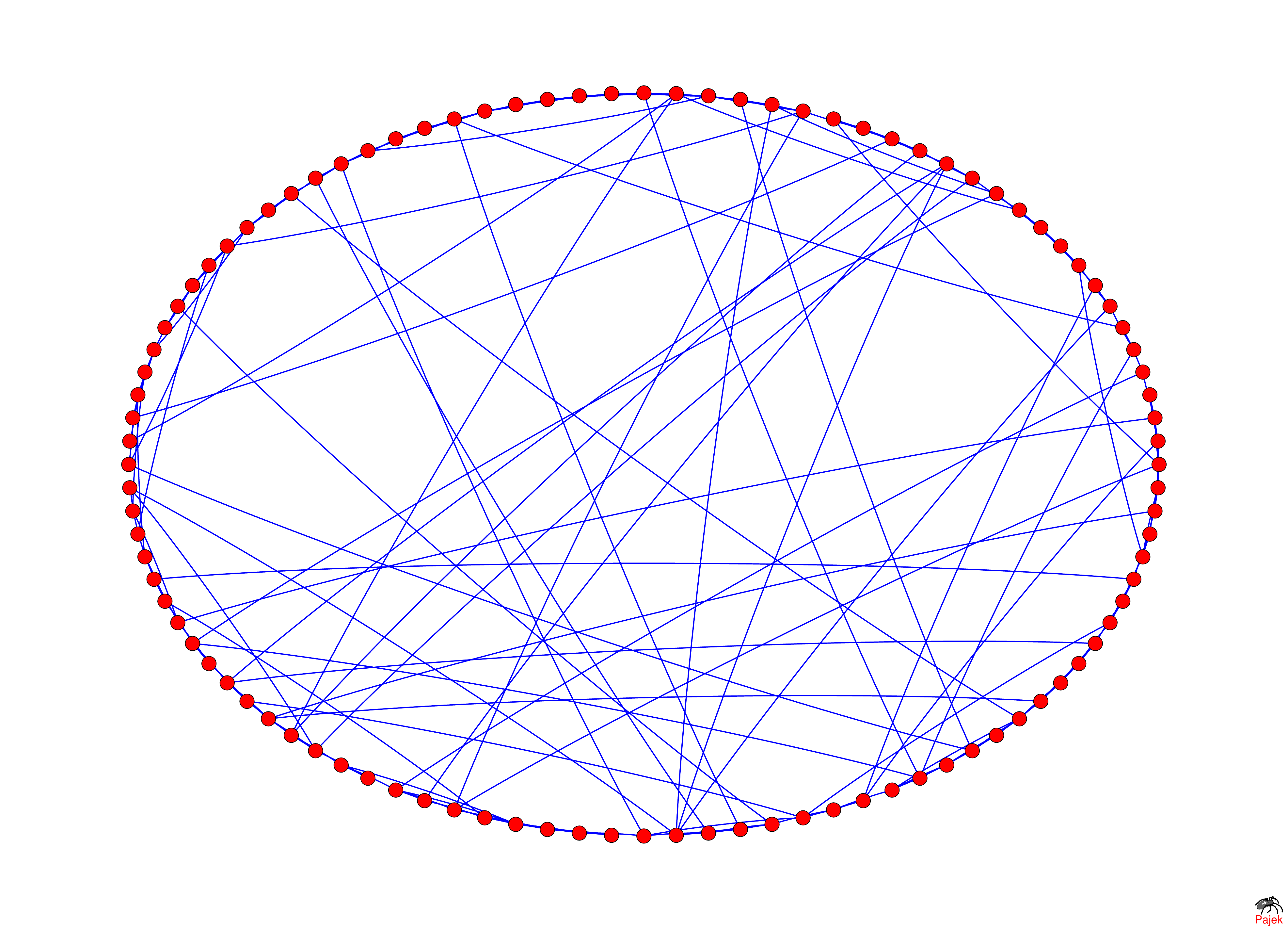}\\
\includegraphics[width=0.36\textwidth]{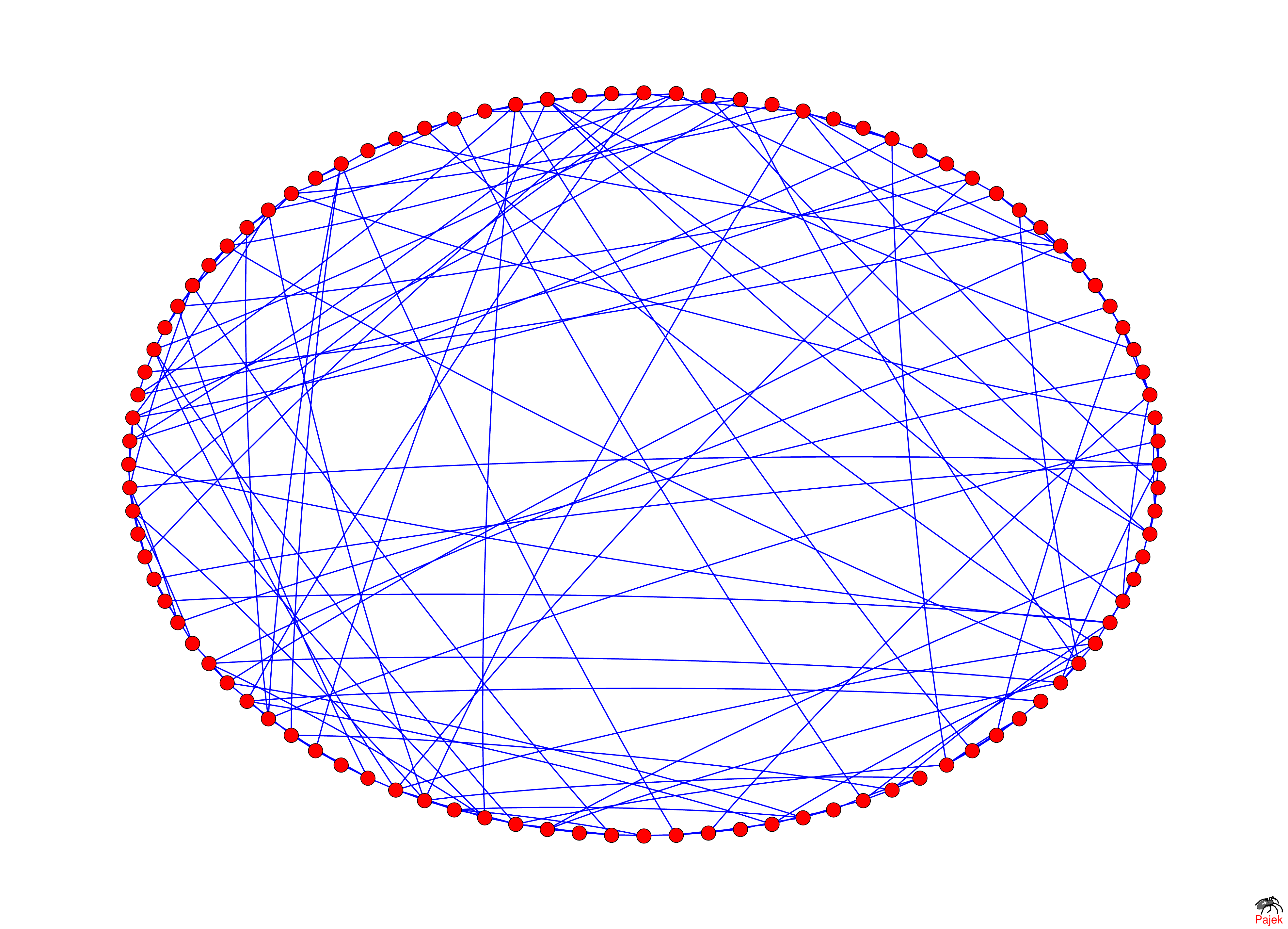}
\end{center}
\caption{\label{fig-net} Sketch of network topology for various clustering coefficients $C_i$. 
$C_i=0.5$, 0.4, 0.3, 0.2, 0.1 from top to bottom, respectively.
The figures were generated with Pajek software \cite{pajek}.}
\end{figure}

For each node $i$ the links to sites $i+1$ and $i+2$ are rewired to two randomly selected nodes so far not-connected to node $i$.
The rewiring procedure occurs with probability $q$.
The examples of original and rewired network are presented in Fig.~\ref{fig-net}.
Increasing rewiring probability $q$ influence the nodes degree distribution and the network clusterization coefficient.
Please note however, that rewiring procedure does not change average nodes degree, {\em i.e.} $\langle k\rangle=N^{-1}\sum_{i=1}^Nk_i=4$.

For given values of rewiring probability $q$ the set $\mathcal{N}(q)=\{\mathcal{N}_1, \mathcal{N}_2, \cdots, \mathcal{N}_M \}$ of $M$ networks is generated.
The clusterization coefficient $C_j$ for $j$-th network is defined as the average over nodes $i=1, \cdots, N$ of the local coefficient $c_i$, where 
\begin{equation}
c_i = \frac{2 y_i}{k_i(k_i-1)}, 
\label{eq:C}
\end{equation}
and $k_i$ is the degree of $i$-th node, {\em i.e.} the number of nodes linked to $i$, and $y_i$ is the actual number of links between these $k_i$ nodes \cite{11}.
The clusterization coefficients $C_j$ for each networks in set $\mathcal{N}(q)$ do not differ more than $\delta C=0.01$ from average values $\mathcal{C}(q)=M^{-1}\sum_{j=1}^{M} C_j$.

The clusterization coefficient for unrewired $(q=0)$ network is exactly equal to $C_j(q=0)=\frac{1}{2}$.

\subsection{\label{sec-process}Contact process description}

The network's nodes are decorated with spin-like variable $s_i\in \{S,D\}$.
Initially (for $t=0$) the $S$ value is randomly assigned to the fraction of $n_S^0\equiv n_S(t=0)$ nodes.
The remaining $(1-n_S^0)N$ nodes are assumed to be in $D$ state.

Every time step ($1\le t\le T$) the random sequence of $N$ nodes' labels is created by sampling with replacement.
Now, network vertices are visited accordingly to this list.

\begin{figure}
\begin{center}
\subfigure[\label{fig-rules-a}]{\includegraphics[width=0.49\columnwidth]{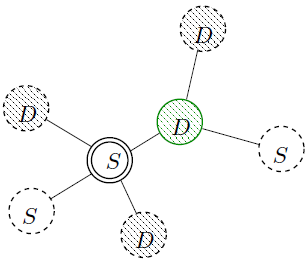}} 
\subfigure[\label{fig-rules-b}]{\includegraphics[width=0.49\columnwidth]{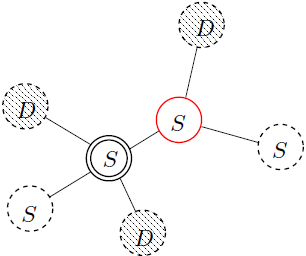}}\\
\rule{0.85\columnwidth}{0.4pt}\\[2mm]
\subfigure[\label{fig-rules-c}]{\includegraphics[width=0.49\columnwidth]{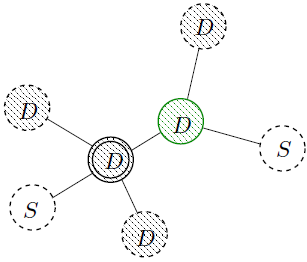}} 
\subfigure[\label{fig-rules-d}]{\includegraphics[width=0.49\columnwidth]{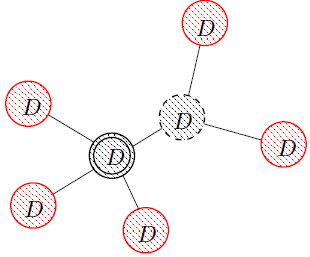}}
\caption{\label{fig-rules} Sketch of contact rules.
(a) The site $S$ influence its nearest-neighbor being in state $D$.
(b) As a result the neighbor is converted to $S$ state.
A pair of connected nodes in $D$ state (c) influence all pair's neighbors which are changed to $D$ (d).
The latter occurs with probability $p$.}
\end{center}
\end{figure}

If the visited node (denoted with double ring in Fig.~\ref{fig-rules}) is marked 
\begin{itemize}
\item as $S$ and at least one of its neighbors is in $D$ state [Fig.~\ref{fig-rules-a}] then the state of this $D$ node is changed to $S$ [Fig.~\ref{fig-rules-b}],
\item as $D$ and at least one of its neighbors is in $D$ state [Fig.~\ref{fig-rules-c}] then with the probability $p$ the state of all the nearest-neighbors of this pair is changed to $D$ [Fig.~\ref{fig-rules-d}].
\end{itemize}

The simulation time $T$ should be long enough to ensure reaching stationary state, {\em i.e.} $dn_S(t\to T)/dt=0$.

\section{\label{sec-results}Results}

\begin{figure}
\begin{center}
\subfigure[$\mathcal{C}=0.1, p=0.3$\label{fig-time-a}]{\includegraphics[width=0.77\columnwidth]{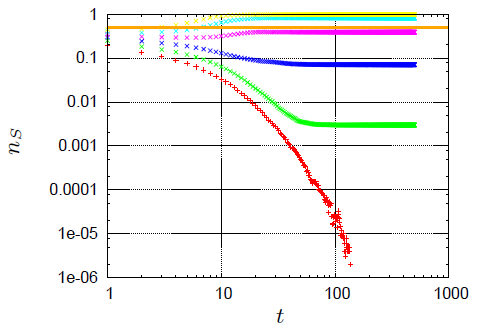}}
\subfigure[$\mathcal{C}=0.2, p=0.4$\label{fig-time-b}]{\includegraphics[width=0.77\columnwidth]{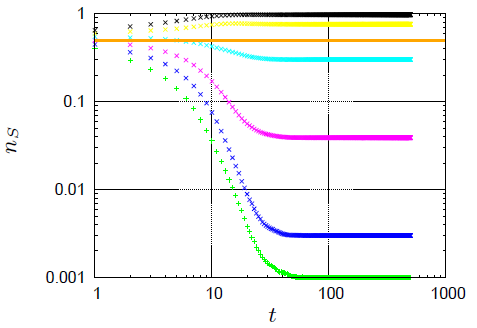}}
\subfigure[$\mathcal{C}=0.4, p=0.6$\label{fig-time-c}]{\includegraphics[width=0.77\columnwidth]{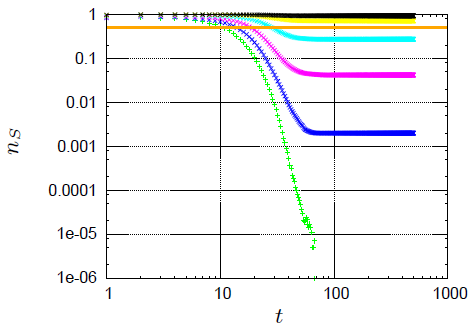}}
\subfigure[$\mathcal{C}=0.5,p=0.41$\label{fig-time-d}]{\includegraphics[width=0.77\columnwidth]{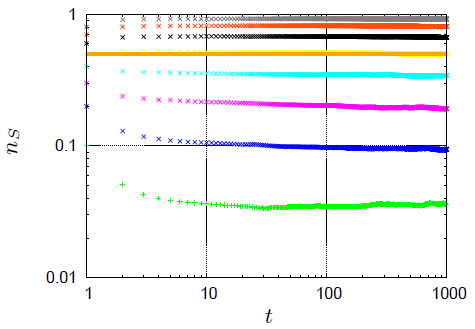}}
\end{center}
\caption{\label{fig-time} Examples of the time evolution of nodes $S$ density $n_S(t)$ for various initial concentrations $n_S^0$ and several set of $(\mathcal{C}, p)$.
The sub-figures correspond to $(\mathcal{C}, p)$ pairs equal to $(0.1, 0.3)$, $(0.2, 0.4)$, $(0.4, 0.6)$ and $(0.5, 0.41)$.
The orange horizontal line indicates $n_S=\frac{1}{2}$.}
\end{figure}

In Fig.~\ref{fig-time} the time evolution of fraction of $S$ nodes $n_S(t)$ are presented.
For a given set of $(\mathcal{C}, p, n_S^0)$ parameters the results of simulation of the contact process described in Sec.~\ref{sec-process} are averaged over $M=10^3$ networks realizations.
These networks differ both in their topology and initial distribution of $S$ nodes.
Please note however, that for unrewired network ($q=0$, $\mathcal{C}=1/2$) only initial distribution of $S$ nodes allows for distinguishing among networks.

\begin{figure}
\begin{center}
\subfigure[$\mathcal{C}=0.2, p=0.4$\label{fig-finalvsini-a}]{\includegraphics[width=0.77\columnwidth]{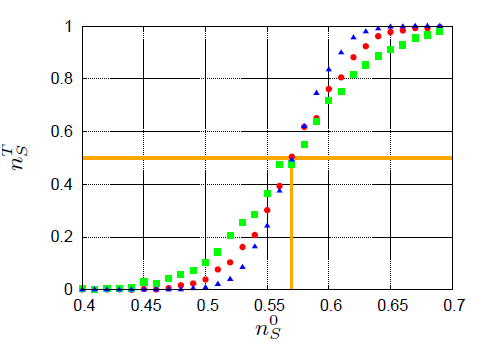}}
\subfigure[$\mathcal{C}=0.3, p=0.5$\label{fig-finalvsini-b}]{\includegraphics[width=0.77\columnwidth]{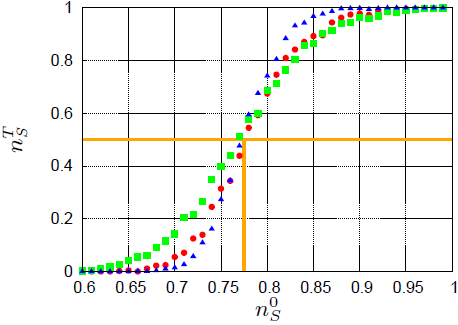}}
\subfigure[$\mathcal{C}=0.4, p=0.4$\label{fig-finalvsini-c}]{\includegraphics[width=0.77\columnwidth]{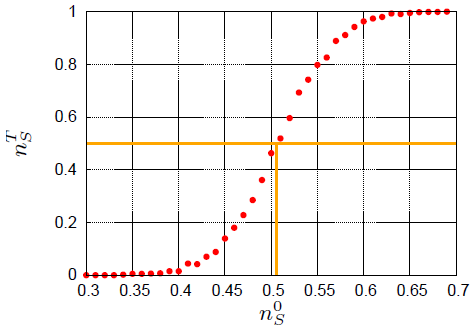}}
\subfigure[$\mathcal{C}=0.5,p=0.41$\label{fig-finalvsini-d}]{\includegraphics[width=0.77\columnwidth]{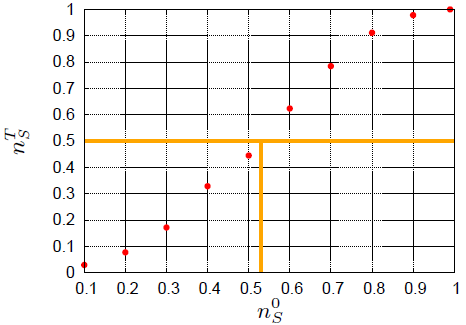}}
\end{center}
\caption{\label{fig-finalvsini} Examples of the dependence of the nodes $S$ final density $n_S^T$ on initial nodes $S$ fraction $n_S^0$.
The curves correspond to various network sizes $N=500$ (\FilledSmallSquare), $1000$ (\FilledSmallCircle), $2000$ (\FilledSmallTriangleUp).
The sub-figures correspond to $(\mathcal{C}, p)$ pairs equal to $(0.2, 0.4)$, $(0.3, 0.5)$, $(0.4, 0.4)$ and $(0.5, 0.41)$.} 
\end{figure}

In Fig.~\ref{fig-finalvsini} the dependence of the nodes $S$ final density $n_S^T\equiv n_S(t=T)$ on initial nodes $S$ fraction $n_S^0$ are presented.
The curves in Fig.~\ref{fig-finalvsini-a} and \ref{fig-finalvsini-b} correspond to various network sizes $N=500$, 1000, 2000.
With enlarging the system size $N$ we expect that these curves become more steeper and steeper tending to Heaviside's function 
\[ n_s^T(n_S^0) \asymp H(n_S^0-n_S^*) \]
in thermodynamical limit, {\em i.e.} for $N\to\infty$.
The common cross point for these curves $n_S^*$ indicate the (unstable) fixed point splitting $n_S^0$ parameter space into two regions:
\begin{itemize}
\item for $n_S^0<n_S^*$ the system evolves more often towards a final state with all nodes in $D$-state [$n_S^T=0$] 
\item while for $n_S^0>n_S^*$ the systems prefers reaching final state with all nodes in $S$-state [$n_S^T=1$].
\end{itemize}
Please note, that ordinates of these points are equal to $n_S^T(n_S^*)\approx\frac{1}{2}$.
This yields a convenient way for rough estimation of abscissas of fixed point basing only on $n_S^T$ vs. $n_S^0$ dependence for single network size (here $N=10^3$).

\begin{figure}
\begin{center}
\includegraphics[width=0.99\columnwidth]{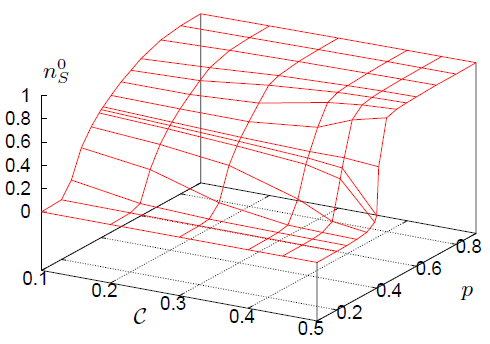} 
\end{center}
\caption{\label{fig-3D} The surface of the unstable fixed points in $(\mathcal{C}, p, n_S^0)$ space.} 
\end{figure}

The surface of the unstable fixed points in $(\mathcal{C}, p, n_S^0)$ space is presented in Fig.~\ref{fig-3D}.
The system evolves more likely towards $n_S^T=1$ for $(\mathcal{C}, p, n_S^0)$ points situated above this surface while points below this surface lead the system more often to $n_S^T=0$.
The points on this surface correspond to such value of initial fraction $n_S^*$ of $S$ nodes (for fixed values $\mathcal{C}$ and $p$) for which their final density is $n_S^T=\frac{1}{2}$.
Of course, reaching the final concentration $n_S^T$ of $S$ nodes exactly equal to $\frac{1}{2}$ is rather rare.
Thus, we estimate $n_S^*$ as
\begin{equation}
n_S^*\approx \frac{n_S^-(0)\left[\frac{1}{2}-n_S^+(T)\right]+n_S^+(0)\left[n_S^-(T)-\frac{1}{2}\right]}{n_S^-(T)-n_S^+(T)},
\end{equation}
where $n_S^{\pm}(T)$ are the values of $n_S^T$ closest to $\frac{1}{2}$ and obeying inequality $n_S^-(T)<\frac{1}{2}<n_S^+(T)$ while $n_S^{\pm}(0)$ are corresponding initial concentration of $S$ nodes leading to these values $n_S^{\pm}(T)$.

\section{\label{sec-summary}Discussion}
Our numerical results indicate that the time evolution drives the system to a homogeneous state where all nodes belong  either to $S$- or $D$-state.
The boundary between the basins of attraction is a surface in the three-dimensional space of parameters: the clustering coefficient $\mathcal{C}$, the probability $p$ of the $D$-process, and the initial concentration of the $S$-nodes.
The boundary consists of unstable fixed points.
The data shown in Fig.~\ref{fig-3D} indicate, that the transition between two homogeneous states is most sharp for $\mathcal{C}=0.5$, {\em i.e.} for the Watts--Strogatz network without rewiring.
Once the rewiring introduces some local disorder, the movement of the boundary between the $S$-phase and the $D$-phase can be stuck on local configurations, and the related metastable states blur the transition.

The advantage of our method of evaluation of the activity of a contact process by counterbalancing it by another contact process is that we get a stationary state which is not a frozen absorbing state,  but a result of a dynamic equilibrium.
The `another process' plays a role of a scale, which allows to compare different processes; if a new process appears, its comparison with the voter dynamics allows to evaluate its efficiency with respect to all processes which had been previously compared with the voter model. Also, we evade the method of quasistationary state, which limits the statistics to surviving trials \cite{2,12a}.

Some possible applications of pair processes have been listed already in Ref.~\cite{9d}.
Here we want to add one, related to computer viruses \cite{13}.
Namely, the direct competition of the processes considered above can find a counterpart in an algorithm of correcting codes in a network of CP units.
The algorithm could be applied to cure the losses made by destructive viruses, which randomly change data.
Once two neighboring computers compare their versions and find them the same, they can safely share this version with all units connected to the pair, because it is unlikely that errors are found at the same place. 
Once they find that their versions are different, both units should be treated as unsafe.
Our results indicate, that the effectiveness of the algorithm depends on the network topology, and in particular on the clustering coefficient.

\acknowledgments{The work was partially supported by the Polish Ministry of Science and Higher Education and its grants for Scientific Research and by the \href{http://www.plgrid.pl/en}{PL-Grid Infrastructure}.}


\end{document}